\documentclass[11pt]{article}

\usepackage[preprint]{acl}

\usepackage{times}
\usepackage{latexsym}

\usepackage[T1]{fontenc}

\usepackage[utf8]{inputenc}

\usepackage{microtype}

\usepackage{inconsolata}

\usepackage{graphicx}

\usepackage{booktabs}
\usepackage[table]{xcolor}
\usepackage[most]{tcolorbox}
\usepackage{makecell}
\usepackage{amsmath}
\usepackage{multirow}
\usepackage{xspace}
\usepackage{tabularx}
\usepackage{caption}
\usepackage{subcaption}
\definecolor{sectiongray}{gray}{0.9}

\definecolor{softblue}{RGB}{100, 130, 180}
\definecolor{softblueback}{RGB}{244, 247, 251}
\definecolor{softbluetitle}{RGB}{225, 233, 244}

\definecolor{evgreen}{RGB}{0,110,80}
\definecolor{evred}{RGB}{178,34,34}
\newcommand{\correctclaim}[1]{\textcolor{evgreen}{#1}}
\newcommand{\wrongclaim}[1]{\textcolor{evred}{#1}}

\newtcolorbox{promptbox}[1]{
  enhanced,
  colback=softblueback,
  colframe=softblue,    colbacktitle=softblue,
  coltitle=white,
  fonttitle=\bfseries\small,
  title={#1},
  boxrule=0.8pt,
  arc=2.5mm,
  left=9pt, right=9pt, top=7pt, bottom=8pt}

 \newcommand{\ours}{\textsc{C}o\textsc{ttage}\xspace}
 
\title{\ours: An Explainable Judge for Text-to-Audio Generation}

\author{
  \textbf{Yu-Xiang Lin\textsuperscript{1}},
  \textbf{Yun-Man Hsu\textsuperscript{1}},
  \textbf{Bo-Ru Lu\textsuperscript{2}}%
    \thanks{This work is not related to their position at Amazon.},
  \textbf{Kuan-Po Huang\textsuperscript{1}},
  \textbf{Wenze Ren\textsuperscript{1}},
  \textbf{Ming-To Chuang\textsuperscript{1}},
\\
  \textbf{Chun-Yi Kuan\textsuperscript{1}},
  \textbf{Jiawei Du\textsuperscript{1}},
  \textbf{Byeonggeun Kim\textsuperscript{2}}\footnotemark[2],
  \textbf{Qingming Tang\textsuperscript{2}}\footnotemark[2],
  \textbf{Chieh-Chi Kao\textsuperscript{2}}\footnotemark[2],
\\
  \textbf{Ankur Gandhe\textsuperscript{2}}\footnotemark[2],
  \textbf{Chao Wang\textsuperscript{2}}\footnotemark[2],
  \textbf{Hung-yi Lee\textsuperscript{1,3}}
\\
\\
  \textsuperscript{1}National Taiwan University \quad
  \textsuperscript{2}Amazon AGI \quad
  \textsuperscript{3}NTU AI-CoRE
\\
\texttt{\{f13942075, hungyilee\}@ntu.edu.tw}
}

\begin{document}
\setcounter{footnote}{1} 
\maketitle
\begin{abstract}
Rapid progress in text-to-audio (TTA) generation demands reliable and explainable automatic evaluation.
However, existing metrics fail to simultaneously deliver high correlation with human judgments and meaningful, audio-grounded explanations, forcing reliance on labor-intensive human evaluation that bottlenecks model development iteration.
To address this, we propose \ours\footnote{The code and data are publicly available at \url{https://github.com/YuXiangLin1234/CoTTAGE}}, a reasoning-based evaluator for text relevance (REL) and audio quality (OVL).
Following a reason-then-predict paradigm, \ours first generates an audio-grounded rationale to diagnose errors before predicting scores.
Supported by over 10K newly collected human OVL annotations, \ours is first trained via supervised fine-tuning and then refined with reinforcement learning using a Thurstone-based preference objective that converts sampled score distributions into pairwise preference probabilities and rewards agreement with rankings derived from human judgments.
Across six benchmarks spanning REL and OVL, \ours achieves state-of-the-art correlation with human judgments while producing human-preferred explanations to accelerate TTA development.
\end{abstract}

\begin{figure}[!t]
\centering
\includegraphics[width=\linewidth]{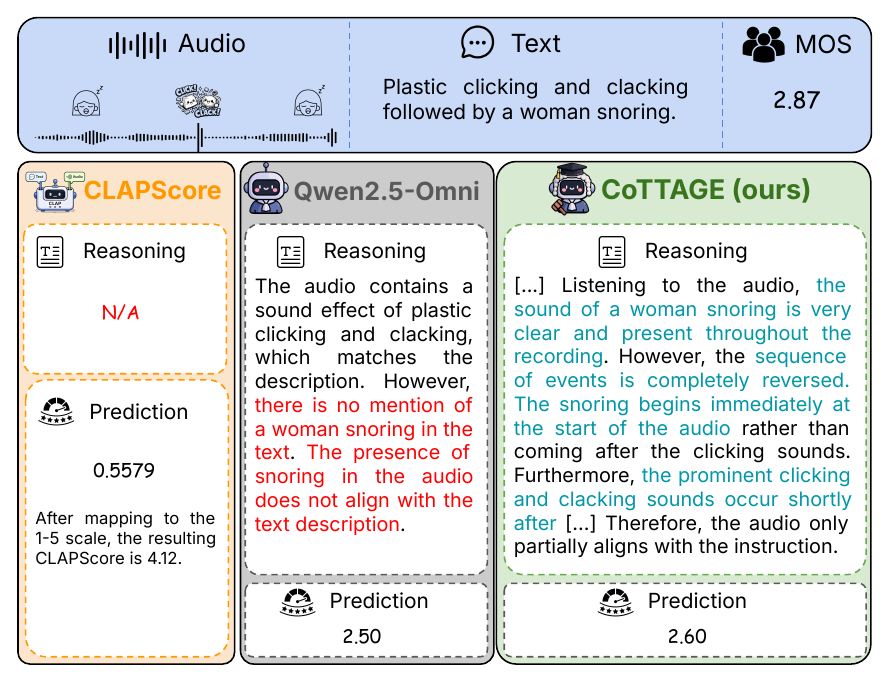}
\caption{
An example of text-audio relevance evaluation with different models.
CLAPScore predicts only a scalar score, while Qwen2.5-Omni often hallucinates reasoning.
In contrast, our \ours model provides an audio-grounded explanation highlighting a temporal-order mismatch.}
\label{fig:reasoning_teaser}
\vspace{-1em}
\end{figure}

\section{Introduction}

Reliable evaluation is fundamental to the development of text-to-audio (TTA) generation systems.
As TTA models continue to improve, effective evaluation becomes increasingly important for guiding model selection and accelerating iterative development~\cite{t2afeedback, ttarl,  ttabench}.
Among existing evaluation approaches, human assessment remains the gold standard, where annotators typically evaluate generated audio from two perspectives: relevance to the text instruction (REL) and overall audio quality (OVL)~\cite{audiogen, audiobox, tango2, impact, huang2026fast}.
While accurate, human evaluation is labor-intensive and difficult to scale, making it impractical for the rapid iteration cycles of TTA development.
This motivates the development of automatic judges that can provide scalable evaluations while maintaining alignment with human perception.

Existing automatic evaluation methods remain insufficient for this purpose.
For REL, CLAPScore~\cite{makeanaudio} measures text-audio alignment using the cosine similarity between pretrained audio-language embeddings, but its correlation with human judgments remains limited~\cite{relate}.
To address this gap, RELATE~\cite{relate} and XACLE~\cite{xacle} introduce human-annotated REL datasets, enabling the development of dedicated REL evaluators.
In contrast, OVL lacks comparable training resources, as publicly available OVL annotations are limited to evaluation-scale datasets~\cite{pam, audiobertscore}.

Beyond accuracy, existing evaluators also provide limited explainability. 
Most methods produce only a scalar score~\cite{makeanaudio, pam, elsa}, offering limited insight for diagnosing model failures and guiding model improvement.
While proprietary frontier audio-language models (e.g., Gemini-3.1 Pro) can generate text explanations, their closed-source nature, high API costs, and privacy constraints render them impractical as scalable, reproducible evaluation benchmarks.
In contrast, fine-grained natural-language feedback has been shown to facilitate both error analysis and model improvement in the text domain~\cite{text2grad}. 
This motivates the need for TTA judges that not only align with human perception but also provide natural-language rationales grounded in the audio.

We introduce \ours, a \textbf{C}hain-\textbf{o}f-\textbf{T}hought \textbf{T}ext-to-\textbf{A}udio \textbf{G}eneration \textbf{E}valuator.
Instead of directly predicting a score, \ours first generates a natural-language rationale grounded in the acoustic evidence before assessing either REL or OVL, as illustrated in Figure~\ref{fig:reasoning_teaser}.
To enable OVL training, we collect human annotations for over 10K audio clips from the RELATE and XACLE training sets, complementing their REL annotations.
\ours is trained in two stages: we first perform supervised fine-tuning (SFT) on human-MOS-guided reasoning traces, and then optimize the model via reinforcement learning (RL) using a Thurstone-based pairwise preference reward~\cite{thurstone} to incorporate prediction uncertainty.

Our contributions are three-fold:
\vspace{-0.4em}
\begin{itemize}\setlength{\topsep}{0pt}\setlength{\itemsep}{1pt}\setlength{\parskip}{0pt}
    \item We propose \ours, a reasoning-based TTA judge that generates audio-grounded rationales for REL and OVL assessment, enabling one model to assess both dimensions. \ours achieves the highest correlation with human judgments across all benchmarks and produces human-preferred rationales.
    \item We collect and release human MOS annotations for OVL on over 10K audio clips from RELATE and XACLE, closing the training data gap for audio quality evaluation.
    \item We show that the proposed Thurstone-based pairwise preference optimization complements pointwise supervised fine-tuning, yielding consistent gains across diverse in-domain and out-of-domain benchmarks.
\end{itemize}

\section{Related Work}

\subsection{Text-to-Audio Evaluation}

Text-to-audio (TTA) systems have advanced rapidly in recent years.
Recent models achieve high-fidelity audio synthesis from free-form text while improving controllability, efficiency, and generation quality~\cite{ audioldm2, ntptta, picoaudio2, tangoflux, resonate, woosh, fdaudio}.

As TTA systems become increasingly capable, reliable evaluation has emerged as a critical challenge. 
PAM~\cite{pam} proposes a reference-free metric based on CLAP~\cite{laionclap} to conduct audio quality assessment.
AudioBERTScore~\cite{audiobertscore} measures embedding similarity between synthesized and reference audio using pretrained audio encoders such as Audio Spectrogram Transformers~\cite{ast}.
ELSA~\cite{elsa} combines global text-audio similarity with event-level alignment via language-queried source separation.
RELATE~\cite{relate} and XACLE~\cite{xacle} advance this line of research by introducing large-scale human-annotated datasets and benchmarks for training and evaluating automatic REL predictors.
Beyond REL and OVL, TTA-Bench~\cite{ttabench} has further broadened the evaluation landscape by incorporating dimensions such as fairness and generation efficiency.
Unlike previous methods, \ours jointly evaluates text relevance and overall audio quality while providing natural-language rationales grounded in the audio.

\subsection{Reasoning-based Audio Judges}

Large audio language models (LALMs), typically comprising an audio encoder~\cite{whisper} and a large language model (LLM) backbone, have recently demonstrated remarkable performance across a wide range of audio understanding tasks~\cite{salmonn, qwen25omni, audioflamingo3, mimoaudio, desta2.5}.
Their strong audio understanding capability has motivated growing interest in using LALMs as automatic judges of audio quality.

Early attempts have primarily focused on speech evaluation. 
QualiSpeech~\cite{qualispeech} first collects human-written justifications and demonstrates that LALMs can be fine-tuned for reasoning-based speech quality assessment. 
Subsequent work extends this direction by exploring instruction-aware evaluation of speaking style~\cite{audiojudge}, improving alignment with human judgments through LLM distillation and preference optimization~\cite{allmasjudge}, and multilingual evaluation with richer task formulations, including pairwise comparison and quality improvement suggestions~\cite{speechllmasjudge}.
These studies demonstrate that LALMs can serve as effective and scalable judges for speech quality.

\section{OVL Annotation}
\label{sec:ovl_collection}

\subsection{Data Collection}
\label{sec:data_collect}
While RELATE~\cite{relate} and XACLE~\cite{xacle} provide training-scale human annotations for REL, publicly available human OVL labels on synthesized audio remain limited to small, evaluation-oriented test sets~\cite{pam,audiobertscore}. 
The closest existing resource is the annotation  used to train Audiobox-Aesthetics~\cite{aesthetics}, which contains approximately 97K clips of audio, each rated by three human annotators. Although this corpus is closely related, several key differences limit its applicability to TTA evaluation. 
First, it has not been publicly released. 
Second, its annotations target a different objective: rather than assigning a single OVL score, annotators rated four aesthetic dimensions, Production Quality (PQ), Production Complexity (PC), Content Enjoyment (CE), and Content Usefulness (CU), which the authors show correlate only moderately with OVL on synthesized sound~\cite{aesthetics}.
Third, it consists exclusively of real-world recordings, introducing a domain mismatch with synthesized audio.

We close this gap by collecting human OVL annotations for 10,362 training clips, including 2,862 clips from RELATE and 7,500 clips from XACLE. 
We recruit annotators through Amazon Mechanical Turk, an online participant recruitment platform, and collect five independent ratings for each clip. 
Details of the annotation protocol, quality assurance procedures, and worker payment are provided in Appendix~\ref{app:human_annotation}.

\begin{figure}[!t]
\centering
\includegraphics[width=\linewidth]{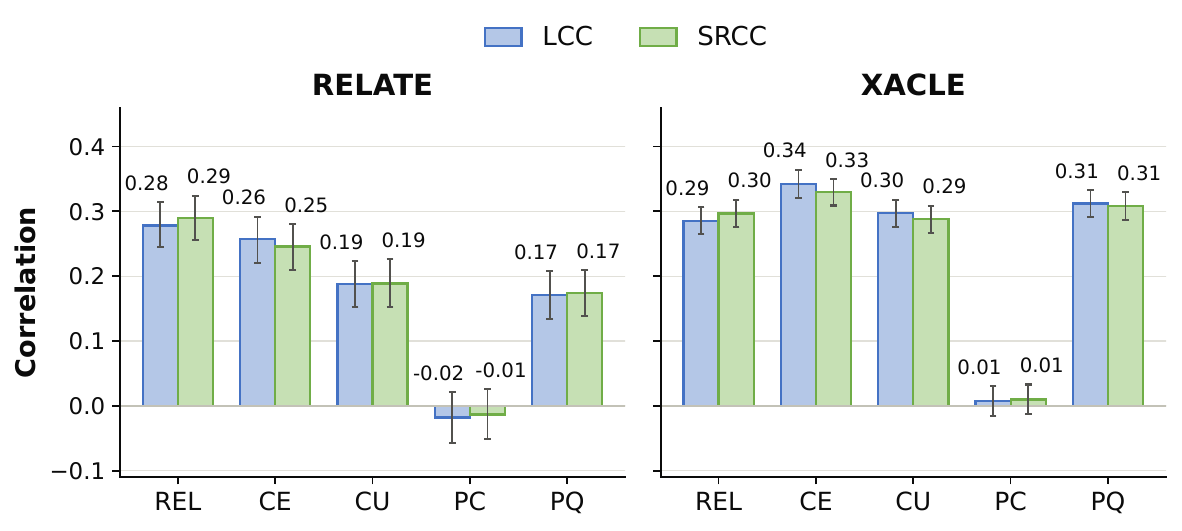}
\caption{Pearson (LCC) and Spearman (SRCC) correlations between human OVL ratings and existing evaluation dimensions, including REL and the four Audiobox-Aesthetics dimensions. 
Error bars indicate bootstrap 95\% confidence intervals.}
\label{fig:ovl_correlation}
\end{figure}

\subsection{Analysis}
\label{ssec:ovl_analysis}

To further justify the need for dedicated OVL annotations, we evaluate how well existing signals predict human OVL ratings.
We consider human REL annotations and the four aesthetic dimensions predicted by the pretrained Audiobox-Aesthetics model~\cite{aesthetics}, since the original human ratings used to train the model are not publicly available.
Figure~\ref{fig:ovl_correlation} reports the Pearson (LCC) and Spearman (SRCC) correlations between each signal and the mean human OVL rating.
The two correlation coefficients closely agree, indicating that the observed trends are not driven by nonlinearity or outliers.

REL shows only weak correlations with OVL on both RELATE and XACLE.
A clip that faithfully matches its prompt is not necessarily perceived as high-quality audio, and vice versa.
The relatively weak correlations indicate that text relevance and overall quality reflect different aspects of human perception.
The four aesthetic dimensions show similarly weak correlations with OVL.
Content Enjoyment is the strongest correlate with OVL, but its correlation is only comparable to that with REL.
Production Complexity is essentially uncorrelated with OVL.
Production Quality, despite being nominally the closest concept to audio quality, reaches only a moderate correlation with OVL on XACLE and performs substantially worse on RELATE.
These results are also consistent with the moderate correlations between individual aesthetic dimensions and OVL reported in~\citet{aesthetics}.

Overall, no existing signal correlates with human OVL beyond LCC $= 0.34$.
Neither REL annotations nor existing aesthetic predictors provide an adequate substitute for dedicated OVL annotations.

\begin{figure*}[t]
    \centering
    \includegraphics[width=\linewidth]{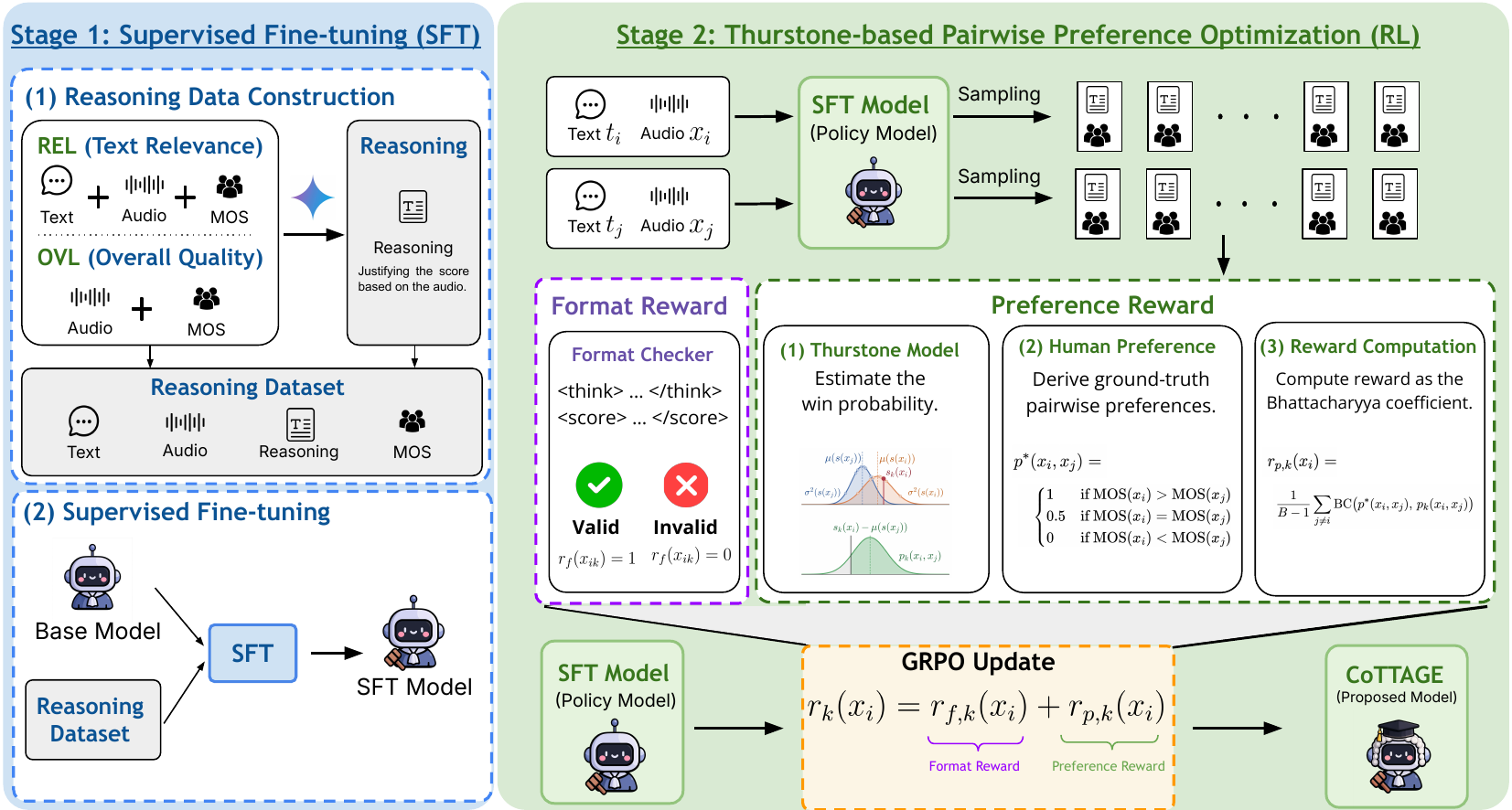}
    \caption{The overall training pipeline of \ours. 
    }
    \label{fig:method_overview}
\end{figure*}

\section{Method}
\label{sec:method}

Figure \ref{fig:method_overview} illustrates the overall training pipeline of \ours.
Our training pipeline consists of two stages: (1) supervised fine-tuning (SFT) to establish foundational reasoning and scoring capabilities (Section~\ref{sec:sft}), and (2) reinforcement learning (RL) with an uncertainty-aware reward to further enhance the model (Section~\ref{sec:rl}).

\subsection{Supervised Fine-Tuning}
\label{sec:sft}
SFT establishes the model's basic reasoning and prediction capabilities.

\paragraph{Reasoning Data Construction.}
Existing general-purpose audio preference datasets provide limited
supervision for reasoning-based judgment.
To obtain reasoning supervision, we generate a reasoning trace for each audio clip while preserving its human-annotated MOS score.
Specifically, we prompt Gemini-3.1 Pro with the audio and the ground-truth MOS score. 
For REL, we additionally provide the corresponding text instruction, since the score depends on text-audio alignment. 
For OVL, following the standard OVL annotation protocol~\cite{audiobox}, no text instruction is provided because the evaluation concerns only the perceptual quality of the generated audio.
The model then generates a rationale explaining how the score is determined by assessing REL and OVL.
Providing the ground-truth score as part of the prompt encourages the generated rationale to remain consistent with the human annotation. 
The instruction we use is shown in Appendix~\ref{app:experimental_setup}.  

\paragraph{Training.}
Using this dataset, we first perform SFT to initialize the model with reasoning and scoring abilities.
During training, the model is required to generate a rationale before predicting the final score. 
This encourages the model to justify its decision through an explicit and human-readable reasoning instead of directly regressing to a score.
After the SFT stage, the model acquires basic reasoning and human preference prediction abilities.

\subsection{Reinforcement Learning}
\label{sec:rl}
Supervised fine-tuning provides only pointwise supervision, where each training sample is associated with a single MOS target. 
While effective for learning an initial evaluator, this formulation offers relatively sparse training signals.
To further improve the judge, we optimize it with reinforcement learning based on pairwise comparisons.
This allows each sample to contribute supervision through comparisons with other samples in the batch.

\paragraph{Thurstone Model.}
We adopt the grouped sampling mechanism of GRPO~\cite{grpo}, where the policy generates $K$ reasoning trajectories for each audio input $x$. 
Each trajectory produces a predicted score $s(x)$, and the resulting set of $K$ scores induces a predictive distribution over each sample, from which we estimate the mean $\mu(s(x))$ and variance $\sigma^2(s(x))$.
The variance reflects the model's inconsistency across reasoning trajectories, with larger values indicating higher uncertainty in the estimated quality score.

Inspired by prior work on image quality assessment~\cite{thurstoneiqa2, thurstoneiqa5, thurstoneiqa6, thurstoneiqa7}, we adopt the Thurstone model~\cite{thurstone} to enable pairwise preference optimization.
It converts uncertainty-aware score distributions into pairwise preference probabilities.
These probabilities are then used as the reward signal for reinforcement learning.
Specifically, the probability that audio $x_i$ is preferred over $x_j$ is defined as:
\begin{equation}
\resizebox{0.85\columnwidth}{!}{$\displaystyle
p_k(x_i, x_j) = \Phi \!\left( \frac{s_k(x_i) - \mu(s(x_j))}{\sqrt{\sigma^2(s(x_i)) + \sigma^2(s(x_j)) + \epsilon}} \right)
$}
\end{equation}
where $\Phi$ is the standard normal cumulative distribution function and $\epsilon$ is a stability constant that prevents the denominator from vanishing.
A larger combined variance naturally yields softer preference probabilities closer to 0.5, resulting in more conservative rewards for uncertain comparisons.

\paragraph{Mapping MOS to Preferences.}
The ground-truth preference $p^*(x_i, x_j)$ is obtained by binarizing the human-annotated MOS into a comparative target:
\vspace{-1em}
\begin{equation}
\resizebox{0.88\columnwidth}{!}{$\displaystyle
p^*(x_i, x_j) = \begin{cases} 1 & \text{if } \text{MOS}(x_i) > \text{MOS}(x_j) \\ 0.5 & \text{if } \text{MOS}(x_i) = \text{MOS}(x_j) \\ 0 & \text{if } \text{MOS}(x_i) < \text{MOS}(x_j) \end{cases}
$}
\end{equation}

\paragraph{Preference Reward.}
Given a batch of $B$ samples, the reward for sample $x_i$ is averaged over all pairwise comparisons with the other $B-1$ samples:
\begin{equation}
\label{eq:preference_reward}
\resizebox{0.85\columnwidth}{!}{$\displaystyle
r_{p,k}(x_i) = \frac{1}{B-1} \sum_{j \neq i} \mathrm{BC}\big(p^*(x_i, x_j),\, p_k(x_i, x_j)\big)
$}
\end{equation}
where $\mathrm{BC}(p, q) = \sqrt{pq} + \sqrt{(1-p)(1-q)}$ is the Bhattacharyya coefficient between the predicted and target Bernoulli preferences, maximized exactly when $p = q$.

\paragraph{Format Reward.}
During exploration, RL-tuned language models often suffer from format collapse, emitting unstructured text that cannot be parsed~\cite{grpo}. 
To enforce the desired reasoning structure, we add a rule-based format reward $r_{f}(x_i)$.
The model earns $1$ only if its output encloses the reasoning trace in \texttt{<think>...</think>} and the final score in \texttt{<score>...</score>}, and $0$ otherwise:
\begin{equation}
r_{f,k}(x_i) = \begin{cases} 1, & \text{required format is satisfied} \\ 0,  & \text{otherwise} \end{cases}
\end{equation}

\paragraph{Overall Reward.}
The final reward for the $k$-th sampled response combines the two terms:
\begin{equation}
r_k(x_i) = r_{p,k}(x_i) + r_{f,k}(x_i).
\end{equation}

\section{Experiments}
\subsection{Experimental Setup}
\paragraph{Datasets.}
We train our judge model using the training sets of RELATE~\cite{relate} and XACLE~\cite{xacle}. 
RELATE contains 2,862 training audio clips spanning both real recordings and TTA-generated samples, while XACLE provides an additional 7,500 training clips. 
Both datasets come with REL scores measuring the semantic alignment between each audio clip and its corresponding text instruction. We additionally collect OVL annotations for both datasets, as described in Section~\ref{sec:ovl_collection}.

We conduct evaluation on the official test sets of RELATE and XACLE to measure in-domain performance. 
To assess generalization, we further evaluate our model on the out-of-domain datasets PAM~\cite{pam} and Clotho~\cite{clotho, audiobertscore}, which differ from our training data in both annotator pools and data collection protocols.

\begin{table*}[t]
\centering
\small
\renewcommand{\arraystretch}{0.95}
    \begin{tabular}{l cc cc cc cc}
    \toprule
    & \multicolumn{2}{c}{XACLE} & \multicolumn{2}{c}{RELATE} & \multicolumn{2}{c}{PAM-Audio} & \multicolumn{2}{c}{Clotho} \\
    \cmidrule(lr){2-3}\cmidrule(lr){4-5}\cmidrule(lr){6-7}\cmidrule(lr){8-9}
    & LCC $\uparrow$ & SRCC $\uparrow$ & LCC $\uparrow$ & SRCC $\uparrow$ & LCC $\uparrow$ & SRCC $\uparrow$ & LCC $\uparrow$ & SRCC $\uparrow$ \\
    \midrule
    \rowcolor[gray]{0.9}\multicolumn{9}{c}{\textit{Embedding-based Methods}} \\
    CLAPScore
      & 0.472 & 0.398
      & 0.437 & 0.411
      & 0.493 & 0.488 
      & 0.396 & 0.378 
    \\
    ELSA
      & 0.398 & 0.371
      & 0.375 & 0.377
      & 0.497 & 0.467 
      & 0.416 & 0.404 \\
    \midrule
    \rowcolor[gray]{0.9}\multicolumn{9}{c}{\textit{Proprietary LALMs}} \\
    Gemini-3.1 Pro
      & 0.551 & 0.489
      & 0.471 & 0.423
      & 0.473 & 0.442 
      & 0.548 & 0.528 \\
    \midrule
    \rowcolor[gray]{0.9}\multicolumn{9}{c}{\textit{Open-Source LALMs}} \\
    Audio Flamingo 3 
      & 0.344 & 0.300
      & 0.215 & 0.193
      & 0.297 & 0.230 
      & 0.376 & 0.339 \\
    Qwen2.5-Omni-7B 
      & 0.162 & 0.153
      & 0.063 & 0.047
      & 0.319 & 0.305
      & 0.334 & 0.301 \\
    Qwen3-Omni-30B 
      & 0.399 & 0.374
      & 0.256 & 0.256
      & 0.373 & 0.353 
      & 0.438 & 0.444 \\
    \midrule
     \ours (ours)
      & \textbf{0.685} & \textbf{0.652}
      & \textbf{0.594} & \textbf{0.591}
      & \textbf{0.657} & \textbf{0.663} 
      & \textbf{0.607} & \textbf{0.602} \\
      \quad \textit{w/o} RL
      & 0.601 & 0.572
      & 0.519 & 0.516
      & 0.606 & 0.590
      & 0.536 & 0.531 \\
        \bottomrule
    \end{tabular}%
\caption{Correlation metrics (LCC and SRCC) between model predictions and human ratings on REL.}
\label{tab:rel-results}
\vspace{-1em}
\end{table*}

\begin{table}[t]
\centering
\small
\renewcommand{\arraystretch}{0.9}
\setlength{\tabcolsep}{4pt}
    \begin{tabular}{l cc cc}
    \toprule
    & \multicolumn{2}{c}{PAM-Audio} & \multicolumn{2}{c}{Clotho} \\
    \cmidrule(lr){2-3}\cmidrule(lr){4-5}
    & LCC $\uparrow$ & SRCC $\uparrow$ & LCC $\uparrow$ & SRCC $\uparrow$ \\
    \midrule
    \rowcolor[gray]{0.9}\multicolumn{5}{c}{\textit{Embedding-based Methods}} \\
    CLAPScore
      & 0.379 & 0.373
      & 0.201 & 0.223 \\
    PAM
      & 0.587 & 0.610
      & 0.445 & 0.451 \\
    ELSA
      & 0.387 & 0.363
      & 0.305 & 0.283 \\
    \midrule 
    \rowcolor[gray]{0.9}\multicolumn{5}{c}
    {\textit{Regression-based Predictors}} \\
    Aesthetics-PQ
      & 0.355 & 0.363
      & 0.465 & 0.450 \\
    Aesthetics-PC
      & 0.092 & 0.127
      & -0.148 & -0.074 \\
    Aesthetics-CE
      & 0.464 & 0.502
      & 0.317 & 0.340 \\
    Aesthetics-CU
      & 0.396 & 0.405
      & 0.487 & 0.471 \\
    \midrule
    \rowcolor[gray]{0.9}\multicolumn{5}{c}
    {\textit{Proprietary LALMs}} \\
    Gemini-3.1 Pro
      & 0.404 & 0.473
      & 0.512 & 0.502 \\
    \midrule
    \rowcolor[gray]{0.9}\multicolumn{5}{c}{\textit{Open-Source LALMs}} \\
    Audio Flamingo 3
      & 0.326 & 0.332
      & 0.218 & 0.225 \\
    Qwen2.5-Omni-7B
      & 0.221 & 0.206
      & 0.271 & 0.329 \\
    Qwen3-Omni-30B
      & 0.429 & 0.424
      & 0.252 & 0.227 \\
    \midrule
    \ours (ours)
      & \textbf{0.611} & \textbf{0.614}
      & \textbf{0.559} & \textbf{0.558} \\
    \quad \textit{w/o} RL
      & 0.529 & 0.525
      & 0.481 & 0.470 \\
    \bottomrule
    \end{tabular}%
\caption{Correlation metrics (LCC and SRCC) between model predictions and human ratings on OVL.}
\label{tab:ovl-results}
\vspace{-1em}
\end{table}

\paragraph{Baselines.}
We compare our method against embedding-based baselines and general-purpose LALMs.
The embedding-based baselines include CLAPScore~\cite{makeanaudio} and ELSA~\cite{elsa}. 
We implement CLAPScore using the pretrained LAION-CLAP model~\cite{laionclap}.
The evaluated LALMs include Gemini-3.1 Pro, Audio Flamingo 3~\cite{audioflamingo3}, Qwen2.5-Omni-7B~\cite{qwen25omni}, and Qwen3-Omni-30B~\cite{qwen3omni}, where Gemini-3.1 Pro is also used to generate the reasoning labels during the SFT stage.
For all LALMs, we prompt the models to first generate their natural-language reasoning before producing a final score.

We further compare against Audiobox-Aesthetics~\cite{aesthetics},  
a specialized audio quality assessment model. 
As discussed in Section~\ref{sec:data_collect}, 
it predicts four aesthetic dimensions (PQ, PC, CE, and CU) rather than a 
single OVL score. 
We therefore report each of the dimensions as a separate baseline for 
OVL prediction.

\paragraph{Implementation Details.}
We initialize our proposed model from Qwen2.5-Omni-7B~\cite{qwen25omni}, and train it with LoRA~\cite{lora} for parameter-efficient fine-tuning. 
We set the LoRA rank to $r = 64$ and the LoRA scaling factor to $\alpha = 128$.
During the SFT stage, the model is trained for 5 epochs with a learning rate of $1 \times 10^{-4}$. 
For the RL stage, we set  the learning rate to $5 \times 10^{-6}$, the KL divergence coefficient to 0.04, and sample $K = 8$ candidate responses per input. 
The overall reward is a weighted sum of the preference reward and the format reward with equal weight.
All prompts and sweep parameters used in our experiments are provided in Appendix~\ref{app:experimental_setup}.
We conduct each experiment with three random seeds and report the median results in the following sections.
Since the REL annotations from RELATE and XACLE are on a 0-10 scale, we linearly map them to a 1-5 scale to align with other benchmarks.

\subsection{Main Results}
Table~\ref{tab:rel-results} and Table~\ref{tab:ovl-results} report the Pearson linear correlation coefficient (LCC) and Spearman rank correlation coefficient (SRCC) between model predictions and human ratings on REL and OVL, where \ours consistently achieves the highest LCC and SRCC across all benchmarks.

\paragraph{\textit{Are pretrained audio-text embeddings enough for TTA evaluation?}} 
CLAPScore and ELSA evaluate TTA outputs through the similarity of pretrained
audio-text embeddings.
While both achieve moderate correlation with human judgments, they consistently lag behind \ours.
On XACLE, our model achieves an LCC of 0.685, compared with 0.472 for CLAPScore and 0.398 for ELSA, with similar gaps observed across the remaining benchmarks. 
For OVL, CLAPScore and ELSA consistently underperform \ours. 
While PAM performs competitively on the PAM-Audio benchmark, \ours\ achieves a slight improvement over PAM on PAM-Audio and demonstrates a larger advantage on Clotho, leading to the strongest overall performance across OVL benchmarks.
These results suggest that pretrained audio-text embedding similarity provides a strong baseline, but reasoning-based evaluation offers more consistent alignment with human judgments across diverse TTA evaluation settings.

\paragraph{\textit{Can general-purpose LALMs serve as reliable judges?}} 
We investigate whether general-purpose LALMs can serve as TTA evaluators without task-specific training.
Our results show that open-source LALMs remain unreliable.
Qwen2.5-Omni, the backbone from which \ours is initialized, achieves an LCC of only 0.063 on RELATE, and even the stronger Qwen3-Omni performs on par with or below embedding-based metrics on most benchmarks.
This indicates that strong audio understanding alone does not necessarily translate into reliable quality assessment.
Overall, Gemini-3.1 Pro is the strongest baseline on REL, ranking first in three of the four benchmarks, and also demonstrates strong performance on OVL benchmarks.
However, \ours consistently outperforms it across all benchmarks, improving the LCC from 0.551 to 0.685 on XACLE (REL) and from 0.404 to 0.611 on PAM-Audio (OVL).

\paragraph{\textit{Do Aesthetic ratings transfer to OVL prediction?}}

We examine whether Audiobox-Aesthetics can serve as a substitute for a reliable OVL judge. 
Despite being trained on nearly ten times more annotated audio than \ours, none of its four aesthetic dimensions correlates strongly with human OVL ratings. 
Besides, the best-performing dimension varies across benchmarks. 
For example, CE achieves the highest correlation on PAM-Audio, whereas CU performs best on Clotho, suggesting that no single aesthetic axis consistently reflects the perceptual quality of synthesized audio.
In contrast, \ours outperforms all four dimensions on PAM-Audio and Clotho.
These findings support our analysis in Section~\ref{sec:ovl_collection}: the mismatch in the prediction target and the training domain limits the applicability of aesthetic ratings to TTA quality evaluation, highlighting the need for collected OVL annotations.

\paragraph{\textit{Does reinforcement learning help beyond SFT?}}
Finally, comparing our full model with its SFT variant (\textit{w/o} RL) shows that reinforcement learning consistently improves both LCC and SRCC across all REL benchmarks and OVL benchmarks.
This demonstrates that the preference-based reward provides complementary supervision beyond supervised fine-tuning alone.
Notably, the SFT variant already achieves strong performance in most cases, demonstrating that supervised fine-tuning provides a strong initialization for reasoning. 
Reinforcement learning further improves upon this initialization, yielding performance beyond what supervised training alone can achieve.

\subsection{Explainable Chain-of-Thought}

\subsubsection{Human Evaluation}
To assess how well the reasoning traces generated by \ours align with the audio content and human perception, we conduct a human preference evaluation.
We randomly sample 50 examples from the PAM test set and collect reasoning traces for REL and OVL from Qwen2.5-Omni-7B, our SFT variant, Gemini-3.1 Pro, and \ours.
Three human annotators independently perform blinded pairwise comparisons between \ours and each baseline, with the presentation order randomized.
For each comparison, annotators select the reasoning trace that better reflects the audio content and human perception, or indicate a tie when neither is preferred.
We report the win rate of \ours (excluding ties), as well as the breakdown of wins, ties, and losses.

\begin{figure}[t]
\centering
\includegraphics[width=\linewidth]{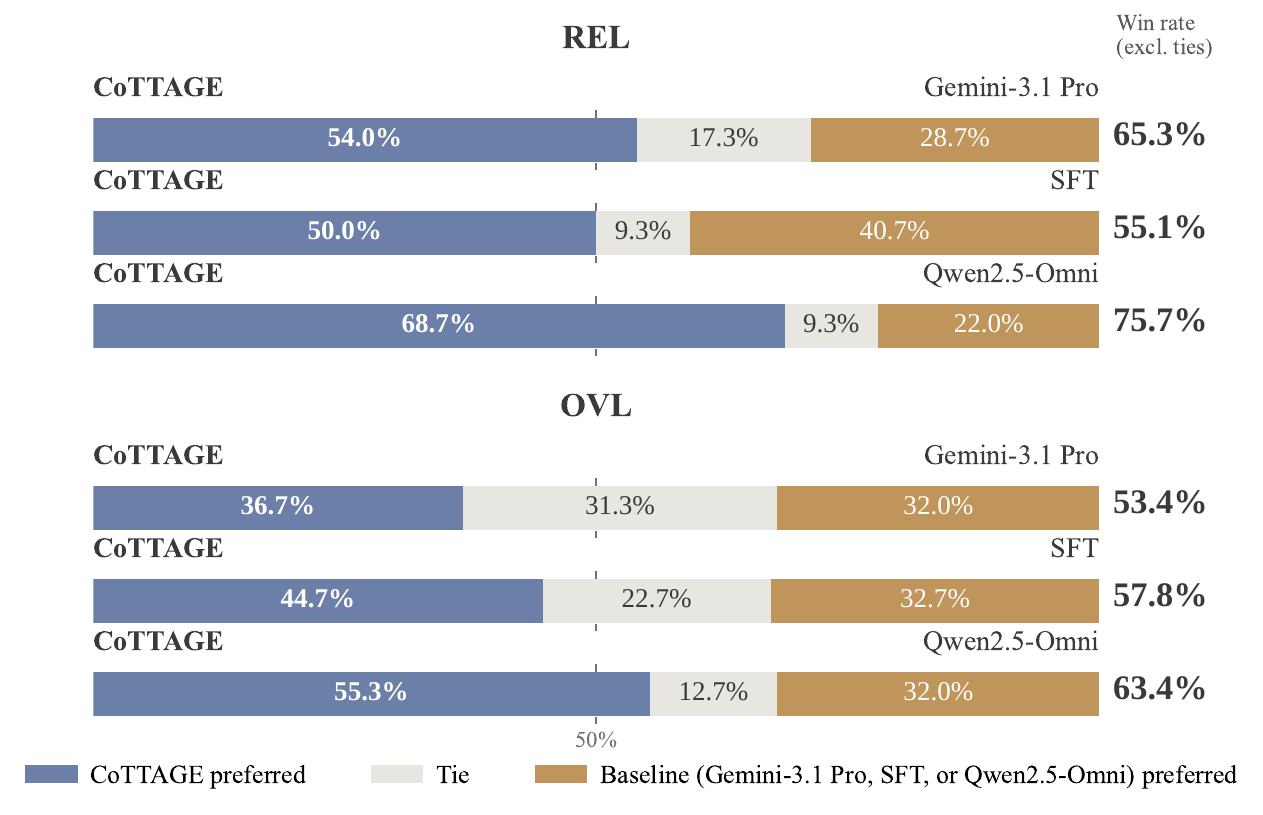}
\caption{Results of human pairwise evaluation of generated scoring rationales. \ours is compared against Qwen2.5-Omni-7B, the SFT-only variant, and the proprietary model Gemini-3.1 Pro.}
\label{fig:reasoning_winrate}
\vspace{-1em}
\end{figure}

Results in Figure~\ref{fig:reasoning_winrate} show that annotators generally prefer \ours over the baselines when ties are excluded.
The largest gains are observed against Qwen2.5-Omni-7B, with win rates of 75.7\% on REL and 63.4\% on OVL.
Compared with the SFT variant, \ours achieves win rates of 55.1\% on REL and 57.8\% on OVL, suggesting that RL further improves the quality of reasoning traces beyond SFT.
The comparison with Gemini-3.1 Pro is particularly competitive, as it is a stronger proprietary model that provides high-quality reasoning traces for distillation and performs well on holistic judgments. Despite this, \ours further improves REL, which requires fine-grained grounding between reasoning traces and audio content.

\begin{table*}[t]
\centering
\scriptsize
\setlength{\tabcolsep}{3.5pt}
\renewcommand{\arraystretch}{0.95}
\begin{tabularx}{\textwidth}{@{}
  c|
  >{\raggedright\arraybackslash}p{1.5cm}
  >{\raggedright\arraybackslash}p{2.5cm}
  >{\hsize=0.9\hsize\raggedright\arraybackslash}X
  >{\hsize=1.1\hsize\raggedright\arraybackslash}X@{}}
\toprule
& \textbf{Type} & \textbf{Instruction} &
\textbf{Qwen2.5-Omni} & \textbf{\ours} \\
\midrule
\multirow{2}{*}[-1.6cm]{\rotatebox[origin=c]{90}{\textbf{REL}}} &
\texttt{Temporal} &
\textit{``A loud motor begins to pick up speed then cuts out and slowly stops.''}\newline {\tiny Human-annotated MOS: 2.10} &
The audio describes \wrongclaim{a motor that starts loudly, then slows down and eventually stops.}
The text also describes \correctclaim{a motor that begins to pick up speed, then cuts out and slowly stops.}~\wrongclaim{The key elements of the motor's behavior (starting loudly, slowing down, and stopping) are consistent between the audio and the text.}
\newline {\tiny Model predicted MOS: 4.50} &
[...] Listening to the audio, there is \correctclaim{a continuous, loud humming or buzzing sound that resembles a motor or drone.}
However, the audio remains \correctclaim{at a constant, steady pitch and intensity from the beginning to the end.}
[...] \correctclaim{there is no sense of the motor ``picking up speed,'' nor is there a sudden ``cutting out'' or stopping sound.}
[...] Therefore, the alignment between the text and the audio is very poor.
\newline {\tiny Model predicted MOS: 1.80} \\
\cmidrule(lr){2-5}

& \texttt{Counting} &
\textit{``A bug chirps once and then starts chirping repetitively.''}\newline {\tiny Human-annotated MOS: 2.00} &
The audio \correctclaim{contains a continuous sound of a bug chirping}, which aligns with the description of a bug chirping repetitively.
\wrongclaim{The description does not mention the bug chirping once before starting to chirp repetitively} [...]
\newline {\tiny Model predicted MOS: 4.50} &
[...] The audio, however, \correctclaim{consists of a continuous, repetitive bug chirping from the very beginning. It completely misses the initial ``chirp once'' action} and plays the repetitive sound throughout its duration. [...]
\newline {\tiny Model predicted MOS: 1.80} \\
\midrule

\multirow{2}{*}[-1.6cm]{\rotatebox[origin=c]{90}{\textbf{OVL}}} &
\texttt{Distortion} &
\textit{``Rapid gunfire with male yelling.''}\newline {\tiny Human-annotated MOS: 1.90} &
[...]~\wrongclaim{The gunshots are clear and distinct, indicating a high-quality sound effect. The man's speech is also clear and understandable,} [...] The overall audio quality is consistent with a real-world recording.
\newline {\tiny Model predicted MOS: 4.50} &
The audio clip \correctclaim{suffers from severe quality degradations throughout its duration.} The prominent sounds, which resemble~\correctclaim{loud bursts or synthetic impacts, are heavily distorted and clipped, resulting in a harsh, abrasive, and unnatural tone.} [...]
\newline {\tiny Model predicted MOS: 2.00} \\
\cmidrule(lr){2-5}

& \texttt{Artifacts} &
\textit{``A jackhammer drilling and vibrating continuously.''}\newline {\tiny Human-annotated MOS: 2.60} &
The audio clip contains \wrongclaim{a sound that resembles a vehicle engine idling}, which is a common and recognizable sound in real-world recordings. The sound is \wrongclaim{clear and consistent, without any noticeable background noise or unnatural patterns.} \wrongclaim{Therefore, the audio quality is excellent.}
\newline {\tiny Model predicted MOS: 5.00} &
[...]~\correctclaim{The most prominent issue with this recording is the significant amount of background noise and static.
A persistent hiss and low-frequency rumble are present throughout the entire clip, which masks the finer details of the mechanical sound.} Additionally, the audio feels somewhat~\correctclaim{muffled and lacks high-frequency clarity}, [...]
\newline {\tiny Model predicted MOS: 2.60} \\
\bottomrule
\end{tabularx}
\caption{Qualitative comparison of natural-language rationales for REL and OVL.
\correctclaim{Green} indicates claims supported by the audio, while \wrongclaim{red} marks unsupported claims.
Rationales are abbreviated.
Full outputs are provided in Appendix~\ref{app:qualitative_results}.}
\label{tab:qualitative}
\vspace{-1em}
\end{table*}

\subsubsection{Qualitative Analysis}
Table~\ref{tab:qualitative} demonstrates examples of rationales generated by \ours and Qwen2.5-Omni.
\paragraph{Reasoning on REL.}
For REL, a consistent failure mode emerges: Qwen2.5-Omni generates rationales that are primarily driven by caption rather than audio. 
Instead of verifying the described events against what it hears, Qwen2.5-Omni tends to restate the caption and then asserts agreement.

In the \texttt{Temporal} example, the audio is actually a steady, unchanging drone. 
Qwen2.5-Omni instead describes a motor that starts loudly, slows down, and eventually stops, matching the caption rather than the audio, and therefore predicts 4.50 for a clip with a human MOS of 2.1.
In contrast, \ours identifies the continuous humming, notes that neither the acceleration nor the final stop occurs, and predicts 1.80.
The \texttt{Counting} example shows the same pattern. 
Although Qwen2.5-Omni recognizes the repetitive chirping, it incorrectly claims that the caption omits the initial isolated chirp and predicts 4.50 for a clip rated 2.0 by annotators. 
\ours instead points out that the repetitive chirping begins immediately and that the isolated chirp never occurs, producing a prediction of 1.80. 

\paragraph{Reasoning on OVL.}
The OVL examples reveal a similar pattern. 
Qwen2.5-Omni frequently produces generic positive rationales such as ``clear and distinct'' and ``without any noticeable background noise or unnatural sound patterns'', assigning high scores even to clips containing obvious synthesis artifacts.
In contrast, \ours identifies the distorted and clipped impulsive sounds throughout the clip and predicts 2.00.

In the~\texttt{Artifacts} example, Qwen2.5-Omni mistakes the jackhammer for an idling engine and still describes the recording as clean, leading to a prediction of 5.00 for a clip rated 2.6 by human annotators. \ours instead identifies the persistent hiss, low-frequency rumble, and muffled timbre, producing a prediction of 2.60.

Overall, these examples demonstrate that \ours achieves more faithful audio understanding, leading to more accurate and reliable TTA judgments.
When Qwen2.5-Omni produces unsupported rationales, its predicted scores are often inaccurate as well.
In contrast, \ours produces audio-grounded rationales that improve explainability  by revealing the evidence behind each prediction, enabling easier diagnosis of TTA model failures and providing insights for improving training data and model design.

\section{Conclusion}

We present \ours, a reasoning-based evaluator that jointly assesses text relevance (REL) and overall quality (OVL) for text-to-audio generation. 
Following the reason-then-predict paradigm, \ours generates a natural-language rationale grounded in acoustic evidence before predicting each score. 
To enable OVL learning, we collect human-annotated OVL labels for over 10K clips from RELATE and XACLE, allowing a single model to assess both dimensions. Building on these resources, we train \ours via SFT on distilled reasoning traces, followed by RL with a Thurstone-based pairwise preference reward. 
Across six benchmarks spanning REL and OVL, \ours achieves the highest correlation with human judgments among all baselines, and its generated rationales are preferred by humans over those of the base model.
We hope that our model and annotations will support further progress in this direction.

\section*{Limitations}

While \ours provides an explainable evaluator aligned with human perception, we outline its key limitations below.

First, \ours relies on human MOS supervision from RELATE and XACLE. Their audio domains, caption styles, and annotator pools may not fully cover broader TTA scenarios, such as music-centric synthesis, speech-centric generation, or non-English instructions. While \ours demonstrates strong transferability to out-of-domain benchmarks such as PAM and Clotho, extending supervision to more diverse audio domains remains an important open direction.

Second, current evaluation benchmarks primarily consist of short audio clips (less than 30 seconds). 
Evaluating long-form audio generation (e.g., multi-minute soundscapes or extended compositions) requires tracking long-range acoustic dynamics and structural continuity over time, which is not fully addressed by short-clip evaluators.

Third, while \ours generates natural-language rationales to diagnose acoustic flaws and event ordering, it does not output exact temporal timestamps to pinpoint when specific acoustic events or artifacts occur. Incorporating explicit temporal grounding represents a promising step toward fine-grained error localization.

\section*{Ethics Statement}

The human OVL annotations in this work were collected via Amazon Mechanical Turk with fair worker compensation aligned with standard crowdsourcing practices. The task involves rating environmental sound clips only, posing minimal risk to participants. Workers were informed that their anonymized ratings would be released for academic research, and no personally identifiable information was collected or stored.

Regarding broader impact, \ours provides scalable and reproducible feedback to guide TTA model development. 
However, directly optimizing a generative model solely against an automatic metric may lead to metric over-fitting. 
Therefore, as with any automatic metric, \ours should be used to complement rather than fully replace periodic human evaluation.

We do not identify any specific harm associated with our model or the collected annotations.

\ours is intended to complement rather than replace human
evaluation, and the annotation task poses minimal risk to participants, from whom no personal information is collected.


\section*{Acknowledgments}

This work was financially supported by Amazon AGI and the Ministry of Education (MOE) of Taiwan under the project Taiwan Centers of Excellence in Artificial Intelligence, through the NTU Artificial Intelligence Center of Research Excellence (NTU AI-CoRE).
We thank the National Center for High-performance Computing (NCHC) of the National Applied Research Laboratories (NARLabs) in Taiwan for providing computational and storage resources.
We also thank Chen-An Li from National Taiwan University for helpful discussions and feedback.

\bibliography{custom}

\appendix

\section{Disclosure of AI Usage}
We used ChatGPT and Claude to polish the manuscript. 
The use of these AI tools did not influence the substantive content, data analysis, or scientific conclusions of the study and served exclusively as a writing aid.
We have carefully checked the AI refinements to ensure that all contents are correct.

\section{Implementation Details}
\label{app:experimental_setup}

\paragraph{Hyperparameter search.}
We sweep the learning rate separately for SFT and RL. 
For SFT, we consider the learning rates
\[
\{1\times10^{-4}, 5\times10^{-5}, 3\times10^{-5}, 2\times10^{-5}, 1\times10^{-5}\},
\]
while for RL, we sweep over
\[
\{1\times10^{-5}, 5\times10^{-6}, 3\times10^{-6}, 2\times10^{-6}, 1\times10^{-6}\}.
\]
For both stages, we sweep the LoRA rank over $\{16,32,64\}$ with $\alpha=2r$.
We additionally sweep the RL sampling temperature over $\{0.8,1.0,1.2\}$.

\paragraph{Prompts.} We provide the full prompts used in our experiments in Figures~\ref{fig:rel-prompt}-\ref{fig:ovl-data-prompt}.
Specifically, Figure~\ref{fig:rel-prompt} and Figure~\ref{fig:ovl-prompt} show the evaluation prompts for REL and OVL, respectively, where the model is instructed to reason step by step before producing a final score between 1.00 and 5.00.
To construct reasoning-augmented training data, we further design the prompts shown in Figure~\ref{fig:rel-data-prompt} and Figure~\ref{fig:ovl-data-prompt}, which condition the model on the ground-truth score and ask it to generate a reasoning trace that naturally leads to this score without explicitly revealing that the
score is given. 
All prompts require the reasoning to be enclosed in \texttt{<think></think>} tags and the final score in \texttt{<score></score>} tags, rounded to two decimal places.

\begin{figure}[t]
\begin{promptbox}{Text Relevance Evaluation Prompt}
Rate the semantic alignment between this audio and the following text description on a scale from 1.00 to 5.00, where 1.00 means no match and 5.00 means a perfect match. Think step by step, then give the final score. Output the score as a number rounded to two decimal places. Put your reasoning inside \texttt{<think></think>} and the final score inside \texttt{<score></score>}.

\medskip
Text: \texttt{\{instruction\}}
\end{promptbox}
\caption{The prompt used for text relevance evaluation.}
\label{fig:rel-prompt}
\end{figure}


\begin{figure}[t]
\begin{promptbox}{Audio Quality Assessment Evaluation Prompt}

Rate the overall audio quality of this audio clip generated by a text-to-audio model on a scale from 1.00 to 5.00, where 1.00 means very poor quality and 5.00 means excellent quality. Think step by step, then give the final score. Output the score as a number rounded to two decimal places. Put your reasoning inside \texttt{<think></think>} and the final score inside \texttt{<score></score>}.
\end{promptbox}
\caption{The prompt used for audio quality assessment.}
\label{fig:ovl-prompt}
\end{figure}


\begin{figure}[t]
\begin{promptbox}{Reasoning Data Construction Prompt - REL}
You are an expert text-to-audio judge. Your job is to evaluate whether an audio faithfully represents the given text instruction.

\medskip
Listen to the audio and compare it with the given text instruction. Carefully analyze how faithfully the audio conveys the content described in the instruction.

\medskip
The ground-truth semantic alignment score has already been determined and must not be changed. Your task is to write a reasoning process that is consistent with this score. The reasoning should read as if you first analyzed the semantic alignment between the instruction and the audio, weighing supporting and conflicting evidence before arriving at a conclusion. Do not explicitly mention that the score is given or explain what the score represents. Instead, let the evidence naturally lead to the final judgment.

\medskip
The semantic alignment score is on a scale from 1.00 to 5.00, where 1.00 means the audio does not match the instruction at all and 5.00 means the audio perfectly and faithfully represents the instruction.

\medskip
Put your reasoning inside \texttt{<think></think>} and output the final score inside \texttt{<score></score>} rounded to two decimal places.

\medskip
Text instruction: \texttt{\{instruction\}}

Ground-truth semantic alignment score: \texttt{\{gt\_score\}}

\medskip
\end{promptbox}
\caption{The prompt used to construct reasoning-augmented training data about REL.}
\label{fig:rel-data-prompt}
\end{figure}


\begin{figure}[t]
\begin{promptbox}{Reasoning Data Construction Prompt - OVL}
You are an expert audio quality judge. Your job is to evaluate the overall quality of an audio clip, i.e., how natural and clean the audio itself sounds, regardless of what it depicts.

\medskip
Listen to the audio and carefully analyze its overall quality, considering aspects such as naturalness, clarity, recording artifacts, distortion, background noise, and any other quality degradations.

\medskip
The ground-truth overall quality score has already been determined and must not be changed. Your task is to write a reasoning process that is consistent with this score. The reasoning should read as if you first analyzed the quality of the audio, weighing supporting and conflicting evidence before arriving at a conclusion. Do not explicitly mention that the score is given or explain what the score represents. Instead, let the evidence naturally lead to the final judgment.

\medskip
The overall quality score is on a scale from 1.00 to 5.00, where 1.00 means very poor quality (heavily distorted, noisy, or unnatural) and 5.00 means excellent quality (clean and natural).

\medskip
Put your reasoning inside~\texttt{<think></think>} and output the final score inside~\texttt{<score></score>} rounded to two decimal places.

\medskip
Ground-truth overall quality score:~\texttt{\{gt\_score\}}

\medskip
\end{promptbox}
\caption{The prompt used to construct reasoning-augmented training data about OVL.}
\label{fig:ovl-data-prompt}
\end{figure}

\begin{figure*}[t]
\centering
\subcaptionbox{Instruction page}{%
  \includegraphics[height=9cm]{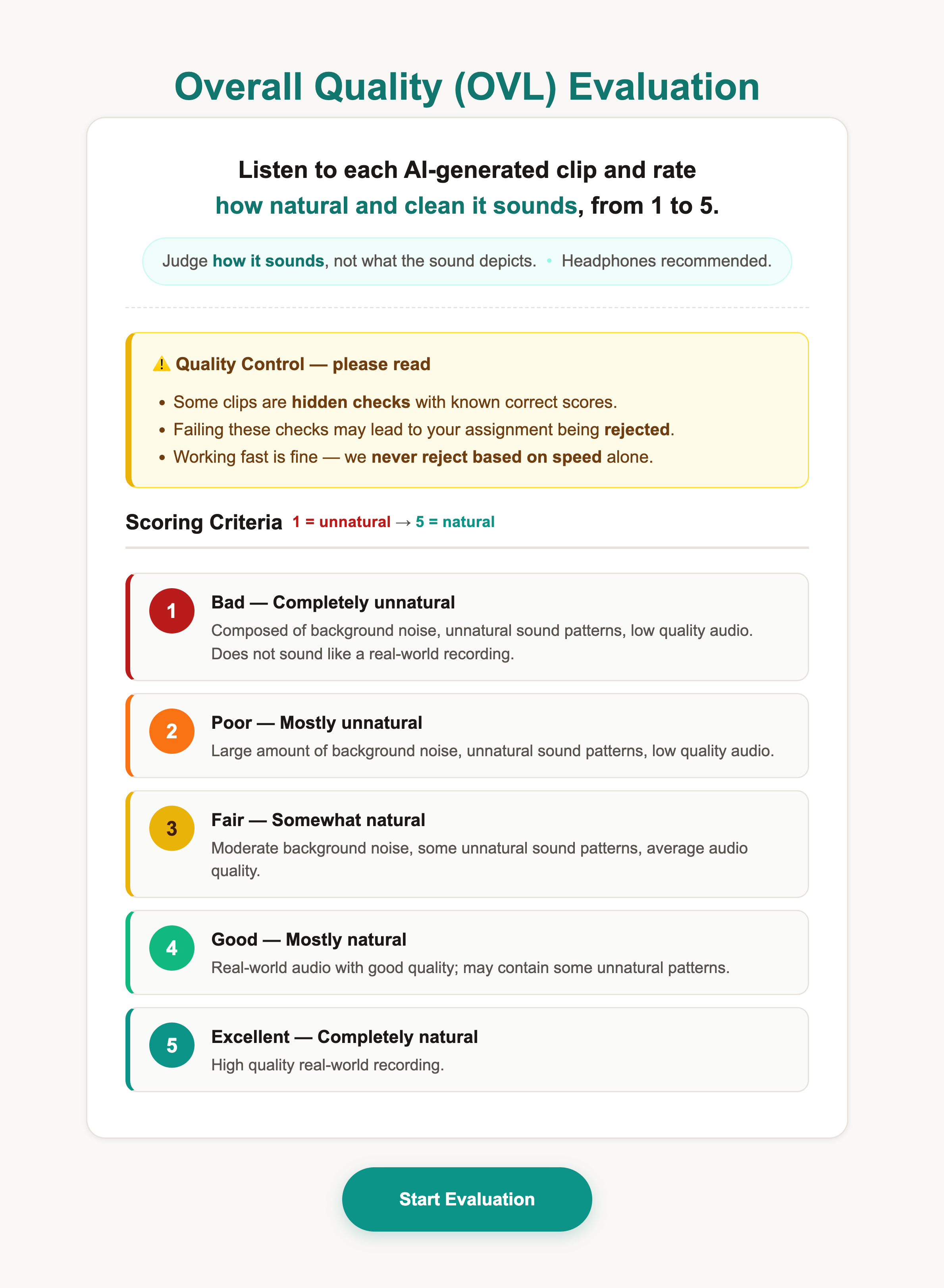}}
\subcaptionbox{Annotation page}{%
  \includegraphics[height=9cm]{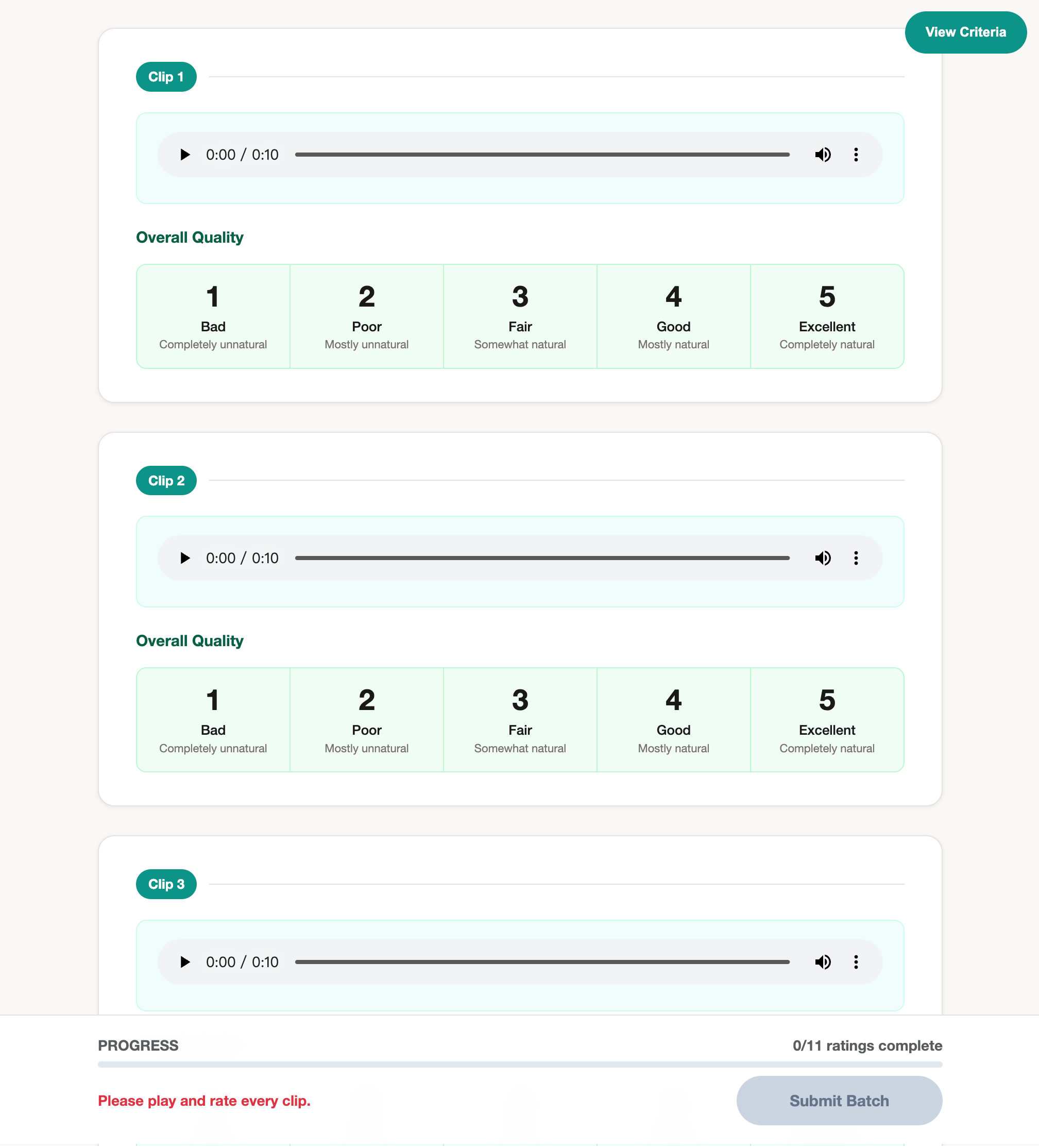}}
\caption{The MTurk annotation interface for OVL collection. For the annotation page, only the first three of the 11 clips are shown.
We adopt the scoring rubric from~\citet{audiobox}.}
\label{fig:ovl_interface}
\end{figure*}

\section{Human Annotation Details}
\label{app:human_annotation}

We collect OVL annotations via Amazon Mechanical Turk (MTurk)\footnote{https://www.mturk.com/} for all audio clips in the training sets of RELATE and XACLE (2{,}862 and 7{,}500 clips, respectively; 10{,}362 in total). 
Each HIT contains 11 audio clips: 10 clips to be annotated and one anchor clip used for quality control, which is randomly drawn from an anchor pool and inserted at a random position so that workers cannot identify it. 
Each HIT is completed independently by five distinct workers. 
To ensure annotation quality, we set the following requirements for annotators:

\begin{itemize}
    \item Annotators must be MTurk Masters, which ensures high-quality workers.
    \item Annotators must have an approval rate higher than 98\%.
    \item Annotators must have more than 10,000 approved tasks.
    \item Annotators must be located in the United States to ensure familiarity with English.
\end{itemize}

Workers rate each clip on an integer scale from 1 to 5, where 1 denotes completely unnatural audio composed of background noise and unnatural sound patterns that does not resemble a real-world recording, and 5 denotes a completely natural, high-quality real-world recording. 
Following the standard OVL annotation protocol \citep{audiobox}, the interface presents only the audio without the corresponding text instruction, and workers are explicitly instructed to judge how the audio \emph{sounds} rather than what it \emph{depicts}. 
The instructions further recommend wearing headphones and inform workers in advance that each HIT contains a hidden quality-control clip and that failing this check may lead to rejection of the assignment.
The final OVL score of each clip is the average of the five worker ratings.
The full worker instructions, scoring rubric, and annotation interface are shown in Figure~\ref{fig:ovl_interface}.

Workers are paid \$0.40 per HIT. To ensure that the payment meets minimum wage standards.
The annotation task involves only rating the quality of machine-generated environmental audio and poses minimal risk to participants.
We do not collect any personal information from workers, and worker identifiers are never released.

\paragraph{Annotation Interface.}
The annotation process consists of two stages.
Upon accepting a HIT, workers first see an instruction page containing the task description, the full scoring rubric for the 1--5 scale, and the quality-control notice.
After reading the instructions, workers click a ``Start Evaluation'' button to proceed to the annotation page.
The annotation page lists the 11 audio clips in order, each accompanied by an audio player and five rating buttons corresponding to scores 1--5.
To prevent workers from forgetting the scoring criteria during annotation, the interface provides a ``View Criteria'' button that reopens the rubric in a pop-up window at any time.
A sticky progress bar at the bottom of the page displays the number of completed ratings in real time.
The interface further requires that every clip be played at least once and rated before the assignment can be submitted, which prevents workers from rating clips without listening to them.
Figures~\ref{fig:ovl_interface}(a) and (b) show the instruction page and the annotation page, respectively.

\paragraph{Quality Control.}
We employ three complementary quality-control mechanisms.
First, we perform anchor-based checking.
We construct an anchor pool consisting of clips with clearly high or clearly low audio quality, each of which is verified through internal listening tests to ensure that its reference score is unambiguous before being added to the pool.
An assignment is rejected if the rating on its anchor clip deviates from the reference score by 2 or more points.
Rejected assignments are republished and re-annotated by other workers, ensuring that every clip eventually receives five valid ratings.
Second, we manage worker access through MTurk qualifications.
Workers with consistently strong performance across batches, i.e., no rejections, high anchor accuracy, and high correlation with the consensus ratings, are granted a trusted qualification and notified early when new batches are published, whereas workers with poor records are assigned an exclusion qualification that blocks them from participating in subsequent batches.
Third, we conduct distribution-based worker monitoring.
On a 5-point scale, a low-effort strategy of compressing all ratings toward the middle of the scale largely evades the anchor-based check, since the resulting anchor errors rarely exceed the rejection threshold.
To detect this behavior, we additionally monitor per-worker rating distributions (the exact-match rate on anchors and the standard deviation of ratings on real clips); workers exhibiting abnormal distributions are manually reviewed and, if confirmed, excluded from subsequent batches via the exclusion qualification.

\begin{table}[t]
\centering
\caption{Inter-annotator agreement of collected OVL annotations, compared with the original REL annotations. We report ICC(1,1), ICC(1,$k$), and Krippendorff's $\alpha$.}
\label{tab:annotation-agreement}
\small
\setlength{\tabcolsep}{4pt}
    \begin{tabular}{l c ccc}
    \toprule
    Dataset & Annotation & ICC(1,1) & ICC(1,$k$) & $\alpha$ \\
    \midrule
    \multirow{2}{*}{RELATE} & REL & 0.449 & 0.762 & 0.446 \\
                            & OVL & 0.317 & 0.699 & 0.317 \\
    \midrule
    \multirow{2}{*}{XACLE}  & REL & 0.435 & 0.755 & 0.435 \\
                            & OVL & 0.322 & 0.704 & 0.322 \\
    \bottomrule
    \end{tabular}%
\end{table}

\paragraph{Inter-Annotator Agreement.}
We report the inter-annotator agreement by ICC(1, 1), ICC(1, k) and Krippendorff's $\alpha$.
We also report the agreement scores of the existing REL annotations for reference. 

As shown in Table~\ref{tab:annotation-agreement}, the ICC(1,$k$) values for RELATE-OVL and XACLE-OVL are 0.699 and 0.704, indicating moderate-to-good reliability under standard guidelines~\cite{agreementmetric1, agreementmetric2}. 
While the ICC(1, 1) and Krippendorff's $\alpha$ values are lower than those reported for the original REL annotations, this difference is expected. 
REL is grounded in a textual prompt and primarily evaluates semantic relevance.
In contrast, OVL requires subjective judgments of overall audio quality and is therefore more susceptible to individual preferences and other confounding factors~\cite{noisymos1, noisymos2}.
Nevertheless, the ICC(1, $k$) values indicate that aggregating ratings across multiple annotators provides sufficiently reliable supervision for training.

\begin{figure*}[!h]
  \centering
  \includegraphics[width=\linewidth]{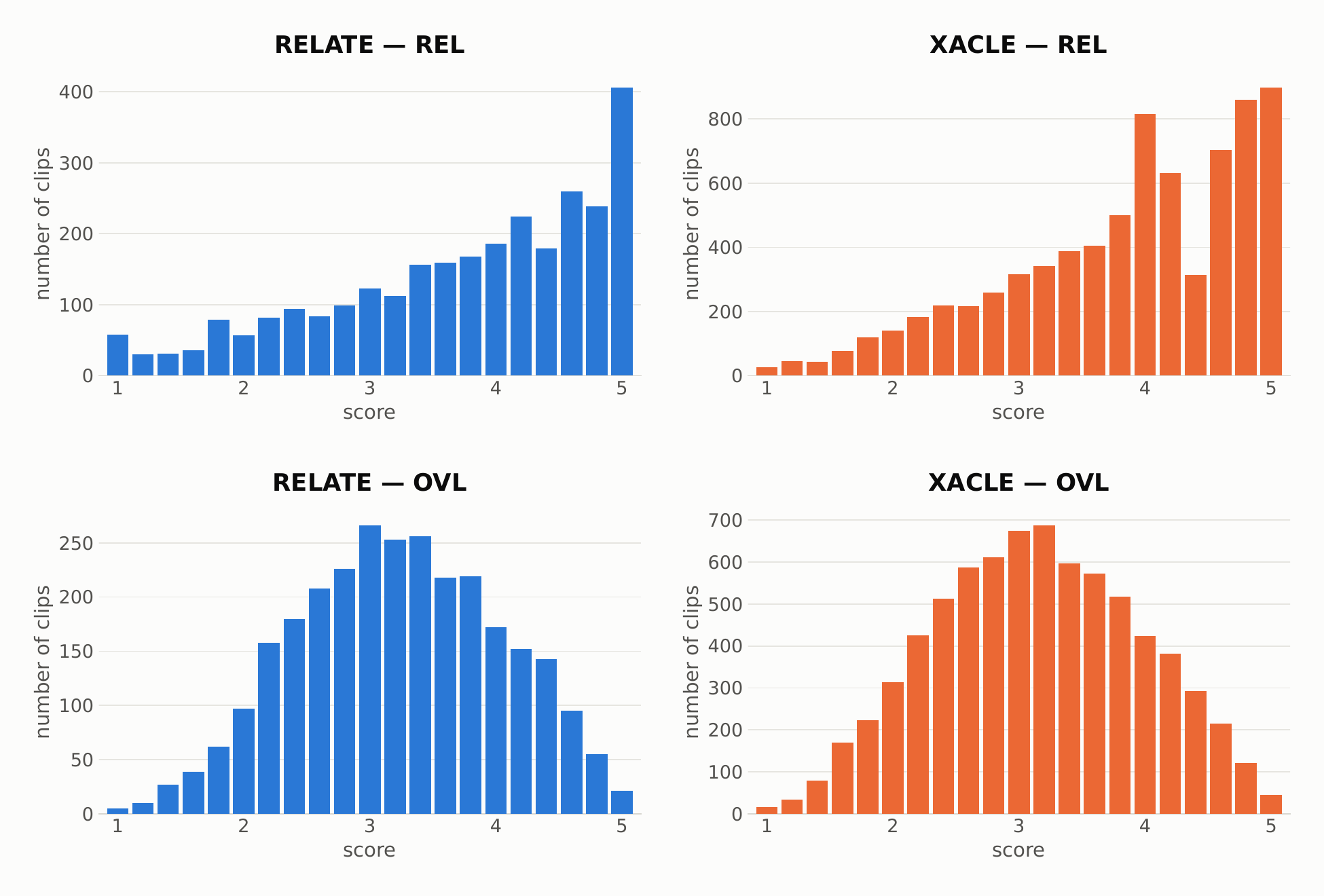}
  \caption{Distributions of the mean human REL and OVL ratings on RELATE and XACLE.}
  \label{fig:score_distribution}
\end{figure*}

\paragraph{Dataset Statistics.}

Figure~\ref{fig:score_distribution} shows the distributions of the mean human REL and OVL ratings
on RELATE and XACLE.
On both datasets, REL is heavily skewed toward the maximum rating of $5$, while OVL is
approximately unimodal and centered around $3$.
Many clips therefore receive near-perfect relevance ratings yet only average quality ratings,
indicating that the two dimensions capture different aspects of human perception.

\section{Qualitative Results}
\label{app:qualitative_results}
We provide qualitative comparisons of the natural-language rationales generated by Qwen2.5-Omni and \ours for REL and OVL in Table~\ref{tab:qualitative_full_rel} and Table~\ref{tab:qualitative_full_ovl}, respectively.

\begin{table*}[h]
\centering
\scriptsize
\setlength{\tabcolsep}{3.5pt}
\renewcommand{\arraystretch}{1}
\begin{tabularx}{\textwidth}{@{}
  c|
  >{\raggedright\arraybackslash}p{1.5cm}
  >{\raggedright\arraybackslash}p{2.5cm}
  >{\hsize=0.9\hsize\raggedright\arraybackslash}X
  >{\hsize=1.1\hsize\raggedright\arraybackslash}X@{}}
\toprule
& \textbf{Type} & \textbf{Instruction} &
\textbf{Qwen2.5-Omni} & \textbf{\ours} \\
\midrule
\multirow{3}{*}[-5.0cm]{\rotatebox[origin=c]{90}{\textbf{REL}}} &
\texttt{Temporal} &
\textit{``A loud motor begins to pick up speed then cuts out and slowly stops.''}\newline {\tiny Human-annotated MOS: 2.10} &
The audio describes \wrongclaim{a motor that starts loudly, then slows down and eventually stops.} The text also describes a motor that begins to pick up speed, then cuts out and slowly stops. \wrongclaim{The key elements of the motor's behavior (starting loudly, slowing down, and stopping) are consistent between the audio and the text.}
\newline {\tiny Model predicted MOS: 4.50} &
The text instruction describes a sequence of events: a loud motor beginning to pick up speed, then cutting out, and finally slowly stopping.

Listening to the audio, there is \correctclaim{a continuous, loud humming or buzzing sound that resembles a motor or drone.} However, the audio remains \correctclaim{at a constant, steady pitch and intensity from the beginning to the end.} It completely lacks the dynamic changes requested in the prompt: \correctclaim{there is no sense of the motor ``picking up speed,'' nor is there a sudden ``cutting out'' or stopping sound.}

While the audio provides a basic motor-like noise, it fails to capture any of the specific actions or the temporal progression detailed in the instruction. Therefore, the alignment between the text and the audio is very poor.
\newline {\tiny Model predicted MOS: 1.80} \\
\cmidrule(lr){2-5}

& \texttt{Counting} &
\textit{``A bug chirps once and then starts chirping repetitively.''}\newline {\tiny Human-annotated MOS: 2.00} &
The audio \correctclaim{contains a continuous sound of a bug chirping}, which aligns with the description of a bug chirping repetitively. \wrongclaim{The description does not mention the bug chirping once before starting to chirp repetitively}, but the continuous chirping is consistent with the description of repetitive chirping. \wrongclaim{Therefore, the audio aligns well with the text description.}
\newline {\tiny Model predicted MOS: 4.50} &
The text instruction specifies a bug chirping once initially, followed by repetitive chirping. The audio, however, \correctclaim{consists of a continuous, repetitive bug chirping from the very beginning. It completely misses the initial ``chirp once'' action} and plays the repetitive sound throughout its duration. Because the temporal structure and the distinct events described in the prompt are absent, the audio fails to faithfully represent the instruction.
\newline {\tiny Model predicted MOS: 1.80} \\
\cmidrule(lr){2-5}

& \texttt{Multi-event} &
\textit{``Ducks quack with blowing wind and faint splashes.''}\newline {\tiny Human-annotated MOS: 2.80} &
The audio contains \correctclaim{the sound of ducks quacking}, which aligns with the text description. \wrongclaim{The presence of wind and faint splashes also suggests an outdoor setting}, which is consistent with the idea of ducks in a natural environment. \wrongclaim{The audio captures the essence of the described scene, making the semantic alignment quite strong.}
\newline {\tiny Model predicted MOS: 4.50} &
The text instruction specifies three distinct elements: ducks quacking, blowing wind, and faint splashes.

Listening to the audio, \correctclaim{the sound of ducks quacking is very prominent and clearly recognizable throughout the clip.} However, the other elements described in the prompt are noticeably absent. \correctclaim{There is no audible sound of blowing wind, nor are there any faint splashes to be heard.}

Because the audio successfully captures the primary subject (the ducks quacking) but completely misses the background environmental sounds (the wind and splashes), it only partially fulfills the text instruction. This results in a moderate alignment score.
\newline {\tiny Model predicted MOS: 2.80} \\
\bottomrule
\end{tabularx}
\caption{Full natural-language rationales for REL, corresponding to Table~\ref{tab:qualitative}, with an additional \texttt{Multi-event} example.
\correctclaim{Green} indicates claims supported by the audio, while \wrongclaim{red} marks unsupported claims.}
\label{tab:qualitative_full_rel}
\end{table*}

\begin{table*}[t]
\centering
\scriptsize
\setlength{\tabcolsep}{3.5pt}
\renewcommand{\arraystretch}{1}
\begin{tabularx}{\textwidth}{@{}
  c|
  >{\raggedright\arraybackslash}p{1.5cm}
  >{\raggedright\arraybackslash}p{2.5cm}
  >{\hsize=0.9\hsize\raggedright\arraybackslash}X
  >{\hsize=1.1\hsize\raggedright\arraybackslash}X@{}}
\toprule
& \textbf{Type} & \textbf{Instruction} &
\textbf{Qwen2.5-Omni} & \textbf{\ours} \\
\midrule
\multirow{3}{*}[-5.0cm]{\rotatebox[origin=c]{90}{\textbf{OVL}}} &
\texttt{Distortion} &
\textit{``Rapid gunfire with male yelling.''}\newline {\tiny Human-annotated MOS: 1.90} &
The audio clip contains \correctclaim{a series of gunshots followed by a man speaking.} \wrongclaim{The gunshots are clear and distinct, indicating a high-quality sound effect. The man's speech is also clear and understandable, suggesting good audio quality.} \wrongclaim{There is no background noise or unnatural sound patterns present.} The overall audio quality is consistent with a real-world recording.
\newline {\tiny Model predicted MOS: 4.50} &
The audio clip \correctclaim{suffers from severe quality degradations throughout its duration.} The prominent sounds, which resemble \correctclaim{loud bursts or synthetic impacts, are heavily distorted and clipped, resulting in a harsh, abrasive, and unnatural tone.} This distortion significantly degrades the clarity and dynamic range of the recording. Additionally, there is a pervasive layer of background noise and static that further muddies the audio. \correctclaim{The human voices heard towards the end are muffled, lack definition, and are difficult to discern clearly against the noisy background.} The combination of intense clipping, high noise levels, and overall lack of fidelity makes the listening experience quite poor. These substantial recording artifacts and the absence of any clean audio elements strongly indicate a very low quality rating.
\newline {\tiny Model predicted MOS: 2.00} \\
\cmidrule(lr){2-5}

& \texttt{Artifacts} &
\textit{``A jackhammer drilling and vibrating continuously.''}\newline {\tiny Human-annotated MOS: 2.60} &
The audio clip contains \wrongclaim{a sound that resembles a vehicle engine idling}, which is a common and recognizable sound in real-world recordings. The sound is \wrongclaim{clear and consistent, without any noticeable background noise or unnatural patterns.} The quality of the audio is high, as it accurately represents the sound of an engine idling. \wrongclaim{Therefore, the audio quality is excellent.}
\newline {\tiny Model predicted MOS: 5.00} &
The audio presents a continuous mechanical sound, resembling a small engine or motor running. \correctclaim{The most prominent issue with this recording is the significant amount of background noise and static. A persistent hiss and low-frequency rumble are present throughout the entire clip, which masks the finer details of the mechanical sound.} Additionally, the audio feels somewhat \correctclaim{muffled and lacks high-frequency clarity}, giving it a slightly distant or enclosed characteristic. While the primary sound source remains identifiable, the heavy presence of noise and the overall lack of crispness severely degrade the listening experience, resulting in a below-average quality rating.
\newline {\tiny Model predicted MOS: 2.60} \\
\cmidrule(lr){2-5}

& \texttt{High quality} &
\textit{``Baby crying then man speaking and humming.''}\newline {\tiny Human-annotated MOS: 4.20} &
The audio clip contains \correctclaim{a baby crying, which is a natural sound.} However, \wrongclaim{the quality of the audio seems to be low, with some background noise and possibly some unnatural sound patterns.} \wrongclaim{The audio does not sound like a high-quality real-world recording.}
\newline {\tiny Model predicted MOS: 2.50} &
The audio presents a baby crying, which is the primary sound source. \correctclaim{The crying itself is captured with good clarity and sounds very natural, without any noticeable distortion, clipping, or digital artifacts.} The vocalizations are distinct and easy to hear. However, \correctclaim{there is a minor amount of background noise or room ambiance present throughout the recording.} While this background noise is not overly intrusive or distracting, it prevents the audio from sounding completely pristine or studio-quality. Overall, the recording is clean and the main subject is well-represented, but the slight background hiss slightly diminishes the overall fidelity.
\newline {\tiny Model predicted MOS: 4.00} \\
\bottomrule
\end{tabularx}
\caption{Full natural-language rationales for OVL, corresponding to Table~\ref{tab:qualitative}, with an additional \texttt{High quality} example.
\correctclaim{Green} indicates claims supported by the audio, while \wrongclaim{red} marks unsupported claims.}
\label{tab:qualitative_full_ovl}
\end{table*}

\end{document}